\documentclass[aps,pre,superscriptaddress,twocolumn,floatfix,showpacs,nofootinbib]{revtex4-1}

\usepackage{amsmath}
\usepackage{amssymb}
\usepackage{graphicx}
\usepackage{mathptmx}

\newcommand\bstrut{\rule[-1.0ex]{0pt}{0pt}}

\newcommand{\tint}{\tau_{\mathrm{int}}}
\newcommand{\e}[1]{\ensuremath{\times 10^{#1}}}
\newcommand{\var}{\mathrm{var}}
\newcommand{\REsq}{R_{\mathrm{E}}^2}
\newcommand{\RE}{R_{\mathrm{E}}}
\newcommand{\RGsq}{R_{\mathrm{G}}^2}
\newcommand{\RG}{R_{\mathrm{G}}}
\newcommand{\RHinv}{R_{\mathrm{H}}^{-1}}
\newcommand{\RH}{R_{\mathrm{H}}}
\newcommand{\DDE}{D_{\mathrm{E}}}
\newcommand{\DDG}{D_{\mathrm{G}}}
\newcommand{\DDH}{D_{\mathrm{H}}}
\newcommand{\DDimp}{D_{\text{imp}}}
\newcommand{\Nmin}{N_{\text{min}}}
\newcommand{\Rimp}{R^2_{\text{imp}}}

\newcommand{\avREsq}{\langle R_{\mathrm{E}}^2 \rangle}
\newcommand{\avRGsq}{\langle R_{\mathrm{G}}^2 \rangle}
\newcommand{\avRHinv}{\langle R_{\mathrm{H}}^{-1} \rangle}
\newcommand{\avRimp}{\langle R^2_{\text{imp}} \rangle }

\newcommand{\aE}{a_{\mathrm{E}}}
\newcommand{\aG}{a_{\mathrm{G}}}
\newcommand{\aH}{a_{\mathrm{H}}}
\newcommand{\bH}{b_{\mathrm{H}}}

\newcommand{\ournu}{0.58759700(40)}
\newcommand{\ourstandardnu}{0.5875970(14)}
\newcommand{\ourrgrh}{1.5803940(45)}
\newcommand{\ourdelta}{0.528(8)}
\newcommand{\ourrerg}{6.253531(10)}
\newcommand{\ourDimproved}{0.3464795(45)}
\newcommand{\ourDE}{1.220322(18)}
\newcommand{\ourDG}{0.1951413(26)}
\newcommand{\ourstandardDE}{1.220345(35)}
\newcommand{\ourstandardDG}{0.1951400(80)}

\newcommand\blfootnote[1]{%
  \begingroup
  \renewcommand\thefootnote{}\footnote{#1}%
  \addtocounter{footnote}{-1}%
  \endgroup
}

\usepackage{hyperref}
\hypersetup{
    pdfauthor={Nathan Clisby and Burkhard Duenweg},
    pdftitle={High-precision estimate of the hydrodynamic radius for self-avoiding walks},
    breaklinks = true, colorlinks = true,
    linkcolor = blue,
    anchorcolor = blue,
    citecolor = blue,
    filecolor = blue,
    urlcolor = blue
    }

\begin{document}

\title{High-precision estimate of the hydrodynamic radius for self-avoiding walks$^\dagger$}

\author{Nathan Clisby$^{\ddagger,~}$}

\email[]{nclisby@swin.edu.au}

\affiliation{School of Mathematics and Statistics, The University of
  Melbourne, Victoria 3010, Australia}

\author{Burkhard D\"unweg}

\affiliation{Max Planck Institute for Polymer Research, Ackermannweg
  10, 55128 Mainz, Germany}

\affiliation{Condensed Matter Physics, TU Darmstadt, Karolinenplatz 5,
  64289 Darmstadt, Germany}

\affiliation{Department of Chemical Engineering, Monash University,
  Clayton, Victoria 3800, Australia}
  

\begin{abstract}
The universal asymptotic amplitude ratio between the gyration radius
and the hydrodynamic radius of self-avoiding walks is estimated by
high-resolution Monte Carlo simulations. By studying chains of length
of up to $N = 2^{25} \approx 34 \times 10^6$ monomers, we find that
the ratio takes the value $\RG/\RH = \ourrgrh$, which is several
orders of magnitude more accurate than the previous state of the
art. This is facilitated by a sampling scheme which is quite
general, and which allows for the efficient estimation of averages of
a large class of observables. The competing corrections to scaling for
the hydrodynamic radius are clearly discernible. We
also find improved estimates for other universal properties that
measure the chain dimension. In particular, a method of
analysis which eliminates the leading correction to scaling 
results in a highly accurate estimate for the Flory exponent 
of $\nu = \ournu$.
\end{abstract}



\maketitle

\section{Introduction}
\label{sec:introduction}

A\blfootnote{\hspace{-0.8em}$^\dagger$~Journal reference: Phys. Rev. E \textbf{94}:052102 (2016), \href{http://link.aps.org/doi/10.1103/PhysRevE.94.052102}{link}.}
few\blfootnote{\hspace{-0.8em}$^\ddagger$ N.C. current affiliation: {\em Department
of Mathematics, Swinburne University of Technology, P.O. Box 218,
Hawthorn, VIC 3122, Australia}} years ago~\cite{clisby_accurate_2010,clisby_efficient_2010}, one
of the present authors demonstrated significant progress in
calculating universal properties of self-avoiding walks
(SAWs)~\cite{madras_self-avoiding_2013} on a lattice, which is the
standard model to describe the static equilibrium properties of
isolated polymer chains in good solvent. These advances were made
possible through the use of a recursive data structure called the
SAW-tree, which allows
for very fast checking of self-overlaps in Monte Carlo (MC)
simulations based upon the pivot algorithm~\cite{lal_monte_1969,
madras_pivot_1988,li_critical_1995,kennedy_faster_2002}, such that
chains with up to $N \approx 34 \times 10^6$ monomers (repeat units)
could be studied. Universal quantities that are accessible include 
critical exponents such as the Flory exponent $\nu =
0.587597(7)$~\cite{clisby_accurate_2010}, which connects the mean polymer
size $R$ with the degree of polymerization $N$ via the scaling law
$R \propto b N^\nu$, where $b$ is the typical monomer size, and 
universal amplitude ratios such as the ratio of two
different ways to define the size of the coil. The two most popular
measures are the mean squared radius of gyration, $\avRGsq$,
and the mean squared end-to-end distance, $\avREsq$.
Denoting the coordinates of the monomers by
$\vec r_i$, $i = 1, \cdots, N$, the corresponding observables 
are defined as
\begin{align}
\label{eq:DefineRG}
    \RGsq & = \frac{1}{2 N^2} \sum_{i,j}
    \left\vert \vec r_i - \vec r_j \right\vert^2
    = \frac{1}{N} \sum_i \left\vert \vec r_i - \vec
    R_{\mathrm{CM}}\right\vert^2 ,
\\
    \vec R_{\mathrm{CM}} & = \frac{1}{N} \sum_i \vec r_i ,
\\
\label{eq:DefineRE}
\REsq & = \left\vert \vec r_N - \vec r_1 \right\vert^2.
\end{align}
In the limit of infinite chain length,
Ref.~\cite{clisby_accurate_2010} found the universal ratio
$\lim_{N \rightarrow \infty} \avREsq / \avRGsq \approx 6.254$.

Besides $\RGsq$ and $\REsq$, the hydrodynamic radius is a third
important measure of the coil dimension which is measured in dynamic
light scattering experiments~\cite{doi_theory_1988}. The inverse
hydrodynamic radius is defined as
\begin{align}
\label{eq:DefineRH}
\RHinv & =  \frac{1}{N^2} \sum_{i \ne j} \frac{1}{r_{ij}} , \\
r_{ij} & =  \left\vert \vec r_i - \vec r_j \right\vert, 
\end{align}
with corresponding mean value $\avRHinv$. This gives rise to another
interesting amplitude ratio, $\avRGsq^{1/2} \avRHinv$, which is a
universal constant in the limit of infinitely long chains that we denote
as $\RG/\RH$ by abuse of notation.  In the present paper we will utilize
the efficient algorithm of
Refs.~\cite{clisby_accurate_2010,clisby_efficient_2010} to accurately
calculate this universal quantity.

Only two recent high-resolution simulation studies have attempted to
calculate the asymptotic ratio $\RG/\RH$ with good accuracy: D\"unweg
et al.~\cite{dunweg_corrections_2002} find a value $\RG/\RH =
1.591(7)$, while Caracciolo et al.~\cite{caracciolo_polymer_2006}
quote $\RG/\RH = 1.581(1)$. These values are compatible with each
other, and also agree nicely with the prediction of
renormalization-group calculations~\cite{schafer_internal_1986},
$\RG/\RH \approx 1.595$.  Mansfield and
Douglas~\cite{mansfield_influence_2010} have recently calculated the
hydrodynamic radius in the infinite-chain length limit.  However, while
we calculate $\RHinv$ according to the definition Eq.~\ref{eq:DefineRH},
they define a related quantity $\RH^*$ (which is an expectation value)
via the Stokes-Einstein relation
\begin{align}
    D &= \frac{k_{\mathrm{B}} T}{6 \pi \eta \RH^*} ,
\end{align}
where $D$ is the translational diffusion coefficient of the molecule
in infinitely diluted solution, $k_{\mathrm{B}}$ is the Boltzmann
constant, $T$ is the
absolute temperature, and $\eta$ is the solvent viscosity. $\RHinv$
according to Eq.~\ref{eq:DefineRH} gives rise to the short-time (or
Kirkwood) approximation to the diffusivity, while the true long-time
value differs somewhat from the Kirkwood
value~\cite{mansfield_influence_2010,liu_translational_2003,
sunthar_dynamic_2006}. Therefore their result is not directly
comparable with ours. It will be shown that the present study has been
able to obtain $\avRHinv$ according to Eq.~\ref{eq:DefineRH} with
substantially increased accuracy, and our estimate, $\RG/\RH =
\ourrgrh$, is again in good agreement with
Refs.~\cite{dunweg_corrections_2002,caracciolo_polymer_2006}.

A crucial aspect of the analysis of MC data is the observation that
such simulations necessarily deal with finite chains of length $N$, while
the above-mentioned values for the universal numbers hold in the
asymptotic limit $N \to \infty$. For this reason, a good understanding
of the finite chain length effects (or corrections to scaling) is
imperative for a correct and meaningful extrapolation. This is
particularly true for the hydrodynamic radius since the corrections to
scaling are very
strong~\cite{dunweg_corrections_2002,caracciolo_polymer_2006,
mansfield_influence_2010}. While for $\avRGsq$ and $\avREsq$ the
corrections are given by~\cite{clisby_accurate_2010}
\begin{align}
\label{eq:ExpansionRG}
\left< \RGsq \right> & = 
\DDG N^{2 \nu} \left(1 +  \aG N^{-\Delta_1} + \cdots \right) , \\
\label{eq:ExpansionRE}
\left< \REsq \right> & = 
\DDE N^{2 \nu} \left(1 +  \aE N^{-\Delta_1} + \cdots \right) ,
\end{align}
where the correction-to-scaling exponent $\Delta_1 \approx
0.53$~\cite{clisby_accurate_2010}, the hydrodynamic radius has an
additional correction of order $N^{- (1 - \nu)}$, with an exponent
that is fairly close to $\Delta_1$, but which will ultimately be
the \emph{dominant} correction:
\begin{align}
    \label{eq:ExpansionRH}
    \left< \RHinv \right> &= \DDH N^{-\nu} \left( 1 + \aH N^{- \Delta_1} +
\bH N^{- (1 - \nu)} + \cdots \right) ;
\end{align}
here $\DDG, \DDE, \DDH, \aG, \aE, \aH, \bH$
are non-universal amplitudes.
The origin of the $N^{- (1 - \nu)}$ term has been discussed in detail
in Ref.~\cite{dunweg_corrections_2002}. These arguments shall not be
repeated here; we rather refer the interested reader to that paper.

It turns out that the Monte Carlo sampling of $\RH^{-1}$ with the
algorithm of Refs.~\cite{clisby_accurate_2010,clisby_efficient_2010}
is somewhat more tricky than one might expect at first glance. The
reason for that problem is intricately related to the underlying
recursive data structure, and it will be outlined in
Sec.~\ref{sec:challenge}. We have found a solution to the problem by
inventing a sampling strategy, which will be elucidated in
Sec.~\ref{sec:sampling}. We then proceed in Sec.~\ref{sec:details} 
by outlining computational details of our study. In
Sec.~\ref{sec:analysis} we analyze our data and present a summary of
results including our estimate for $\RG/\RH$, 
and a much improved estimate for $\nu$ obtained by
eliminating the leading correction to scaling.  Our simulations
are more accurate than those of Ref.~\cite{clisby_accurate_2010} and
hence allow us to also present improved estimates for the universal
amplitude ratio $\lim_{N \rightarrow \infty} \avREsq / \avRGsq = \DDE
/ \DDG$ and $\Delta_1$. Finally, we conclude in
Sec.~\ref{sec:conclusion}.

\section{The computational challenge}
\label{sec:challenge}

For our polymer simulations we utilize the pivot
algorithm~\cite{lal_monte_1969,madras_pivot_1988}, which is the most
powerful known method for sampling self-avoiding walks at fixed
length.  For SAWs on the simple cubic lattice with $N$ monomers, the
probability of a pivot move being successful decays as $N^{-p}$ with
$p \approx 0.11$. The standard hash table
implementation~\cite{madras_pivot_1988} then requires mean CPU time
$O(N)$ to generate an essentially new configuration with respect to
global observables such as $\REsq$.  Recent algorithmic
improvements~\cite{kennedy_faster_2002,clisby_accurate_2010,
clisby_efficient_2010} have further increased the relative advantage
of the pivot algorithm over other methods. We utilize the SAW-tree
data structure of Ref.~\cite{clisby_efficient_2010} which allows us to
perform pivot moves for an $N$-step SAW in mean CPU time $O(\log N)$,
resulting in mean CPU time $O(N^p \log N)$ to generate an essentially
new configuration with respect to global observables.

The main ingredient of this implementation is a binary tree data
structure that recursively decomposes a chain into subchains of
decreasing length, until finally the monomer level is reached. Each
node on the tree stores aggregate information about its respective
subchain, such as the coordinates of its center of mass, its
end-to-end vector, its squared radius of gyration, and, most importantly,
its bounding box (the smallest rectangular parallelepiped aligned with
the lattice that completely encloses the subchain). Each geometric
object within a bounding box is stored not in terms of absolute
coordinates, but rather in terms of coordinates \emph{relative} to the
origin and the orientation of the box. Now, a pivot move will always
mean that a geometric transformation (combination of rotation,
reflection, and translation) is applied to some monomers. Instead of
moving all these monomers individually, the algorithm just moves those
bounding boxes that need to be moved. Some bounding boxes will be big,
some small, but the algorithm will always pick those boxes that are as
big as possible. For example, in the simple case that the algorithm
happens to just move the monomers number $1, 2, \cdots, N/2$, only one
single bounding box, corresponding to these monomers, is being
transformed.  Because of the storing of relative coordinates, all the data
within such a box can be left as-is and do not need to be updated. In
other words, the algorithm always attempts to work at the
highest-possible levels of the tree and to avoid the data-intensive
low levels as much as possible. Furthermore, since the coordinates of
a box are known both from the outside and from the inside, this
information makes it possible to recursively retrieve, starting from
the top, the absolute coordinates of any geometric object if they are
needed.

After a node has been updated, it needs to pass information to its
higher-level node. For example, the end-to-end vector, the center
of mass, and the gyration radius at the higher level will be changed,
and so will be the bounding box. From there this passing will be done
recursively all the way to the very top. However, information-passing
to lower levels is \emph{not} needed, and this is what makes the
method fast. It can thus be shown that the number of nodes that need
to be updated is $O (\log N)$. The check for overlaps can also be done
with average case $O (\log N)$ computational complexity. The crucial
observation is here that if two bounding boxes do not overlap, then
this is also true for all monomers that they contain. Only in case of
box overlap further investigation is needed, and this is again done in
a top-down recursive fashion.

It is also clear that the evaluation of the end-to-end vector and of
the center of mass are compatible with that approach. The
end-to-end vector of a subchain that is decomposed into two
sub-subchains is the sum of the end-to-end vectors of those
sub-subchains, and therefore it is sufficient to pass information
just to the higher-level node. Exactly the same statement holds for
the center of mass, where instead of a sum we have an
appropriately weighted average.

Although the method is slightly less obvious,
the gyration radius may also be calculated in such a recursive
fashion, as a few
lines of straightforward algebra show that the following decomposition
holds:
\begin{align}
\label{eq:decompose_rg}
\nonumber
\RGsq
& = 
\frac{N_1}{N}
    \left\{ R_{\mathrm{G}1}^2 + \left\vert \vec R_{\mathrm{CM}1} - \vec
    R_{\mathrm{CM}} \right\vert^2 \right\}
\\
& + 
\frac{N_2}{N}
    \left\{ R_{\mathrm{G}2}^2 + \left\vert \vec R_{\mathrm{CM}2} - \vec
    R_{\mathrm{CM}} \right\vert^2 \right\} .
\end{align}
Here $\RGsq$ is the squared gyration radius of the subchain with $N$
monomers, while $R_{\mathrm{G}1}^2$ and $R_{\mathrm{G}2}^2$ are the
corresponding squared gyration radii of the two sub-subchains, with
$N_1$ and $N_2$ monomers, respectively, while $\vec R_{\mathrm{CM}}$ is
the center of mass of the subchain, and $\vec R_{\mathrm{CM}1}$, $\vec
R_{\mathrm{CM}2}$ are the corresponding centers of mass of the
sub-subchains.  Thus, Eq.~\ref{eq:decompose_rg} allows us to calculate
the gyration radius recursively as well.

However, the hydrodynamic radius is an observable that cannot
be decomposed into sub-observables of subchains. The reason is that
$\RHinv$ involves interactions between distinct monomers and cannot
be written in a form that involves only one-body terms (meaning that
only sums of the form $\sum_i \cdots$ occur, but not terms of the form
$\sum_{ij} \cdots$, $\sum_{ijk} \cdots$ and the like). In contrast,
$\vec \RE$ and also $\RGsq$ can straightforwardly be written in
such a form.

Therefore, calculating $\RHinv$ is in principle much harder than $\RE$ or
$\RG$, because a recursive evaluation cannot be done. The brute-force
approach, in which one would evaluate the full double sum $\sum_{i \ne
j} r_{ij}^{-1}$ for each generated chain conformation, will obviously
not work: the computational complexity of the sum, if done exactly,
scales as $O(N^2)$ (perhaps with an additional factor of $O(\log N)$
depending on the details of the implementation).  This could be improved
to $O(N)$ if the hydrodynamic radius were evaluated via the fast
multipole method~\cite{greengard_fast_1987}. If we were using the
hash table implementation of the pivot algorithm then this would indeed
be a very effective approach, as the mean CPU time to generate a new SAW
would also be $O(N)$. However, both the naive and fast multipole methods 
would
dominate the mean CPU time required to generate a new SAW for the
SAW-tree implementation of $O(N^p \log N)$.  In other words, evaluation
of the full sum for the hydrodynamic radius would lead to an algorithm
for which nearly all advantages of the SAW-tree implementation would be
lost!

Our simple solution, whose computational complexity is
logarithmic in $N$, shall be outlined in the next section. From the
structure of the method as explained below, it is clear that it can be
applied to any observable that has the form $\sum_i A_1(\vec r_i)$,
$\sum_{ij} A_2(\vec r_i, \vec r_j)$, $\sum_{ijk} A_3(\vec r_i, \vec
r_j, \vec r_k)$, and so on, as well as combinations of these, and is
thus quite general. However, it may fail if one is interested in
complex observables such as knot types.

\section{Sampling strategy for calculation of the hydrodynamic radius}
\label{sec:sampling}

\begin{figure}[b]
  \begin{center}
    \includegraphics[width=0.975\columnwidth]{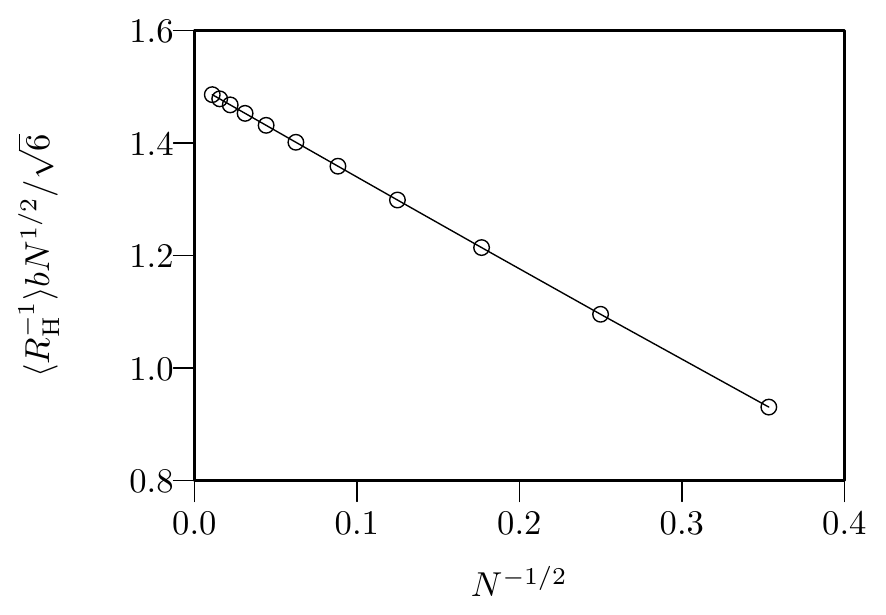}
  \end{center}
  \caption{Exact results for $\avRHinv$ for a Gaussian
    chain. Instead of $\avRHinv$ we rather plot the
    dimensionless ratio $\avRHinv b N^{1/2} /
    \sqrt{6}$, where $b N^{1/2} / \sqrt{6}$ is the asymptotic
    long-chain value for the gyration radius $\avRGsq^{1/2}$. 
    In other words, corrections to scaling are taken
    into account only for the hydrodynamic radius but not for the
    gyration radius. The argument on the abscissa, $N^{-1/2}$,
    reflects the leading-order correction to scaling. Note
    also that the asymptotic value for $N \to \infty$ is
    $\RG / \RH = 8 / (3 \sqrt{\pi}) \approx 1.5045$.}
  \label{fig:GaussianExact}
\end{figure}

The key to our approach to solve the abovementioned problem is the
following simple observation: we write
\begin{align}
\RHinv
    & =
\frac{1}{N^2} \sum_{i \ne j} \frac{1}{r_{ij}}
=
\left(1 - \frac{1}{N} \right) \frac{1}{N (N - 1)}
\sum_{i \ne j} \frac{1}{r_{ij}}
\nonumber
\\
& =
\left(1 - \frac{1}{N} \right) \left[ \frac{1}{r} \right] ,
    \label{eq:sumtoaverage}
\end{align}
where $[ \cdots ]$ denotes an average over all pairs. This
means that, for a given conformation of the chain, we can find the
observable $\RHinv$ not only by brute-force calculation of the sum,
but also by Monte Carlo sampling: we simply pick a pair of monomers
$(i,j)$ uniformly at random from the set of all monomer pairs, and
evaluate $r_{ij}^{-1}$. If we do this often, and average over the
results, this will stochastically converge towards $\RHinv / (1 -
N^{-1})$. Actually, it is sufficient to do this only \emph{once} per
generated chain conformation, since the average over pairs will be
automatically included in the overall sampling. We thus write
\begin{align}
    \left\langle \RH^{-1} \right\rangle &=
  \left( 1 - \frac{1}{N} \right)
  \left\langle \left[ \frac{1}{r} \right] \right\rangle ,
\end{align}
where the average $\langle \cdots \rangle$ means the average over chain
conformations, and $[ \cdots ]$ the average over monomer
pairs; these averaging operations are interchangeable. 

This strategy gives rise to $O(\log N)$ computational complexity for the
operations being done for one chain conformation, since finding the
actual coordinates of monomers $i$ and $j$ involves a recursive search
along the binary tree. In other words, the computational complexity of the
observable evaluation is comparable to the computational complexity to
perform a single update by attempting to perform a pivot move.

In order to test this idea, we first studied a Gaussian chain in
three-dimensional continuous space, with $\langle r_{ij}^2 \rangle =
b^2 \vert i - j \vert$, as a simple toy model. For this model one
finds analytically by a Gaussian integral $\langle r_{ij}^{-1}
\rangle = \sqrt{6 / \pi} b^{-1} \vert i - j \vert^{-1/2}$, and the
remaining double sum is easily numerically evaluated to yield an exact
value for $\avRHinv$ for any reasonable chain length (including all
corrections to scaling). The result is shown in
Fig.~\ref{fig:GaussianExact}.

\begin{figure}[t]
  \begin{center}
    \includegraphics[width=\columnwidth]{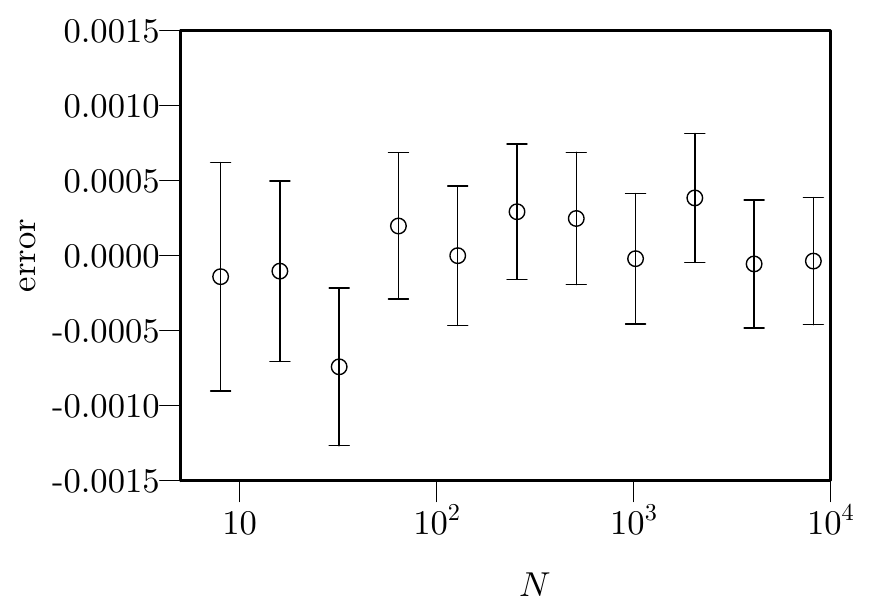}
  \end{center}
  \caption{Difference between $\avRHinv b N^{1/2} / \sqrt{6}$ (sampled
    value) and $\avRHinv b N^{1/2} / \sqrt{6}$ (exact value), as a
    function of chain length $N$. Here the sampled value for
    $\avRHinv$ results from averaging over $10^6$ independent chains,
    using a full evaluation of the double sum $\sum_{i \ne j}
    r_{ij}^{-1}$. The error bars have been estimated as three times
    the standard error of mean.}
  \label{fig:GaussianFullSample}
\end{figure}

\begin{figure}[b]
  \begin{center}
    \includegraphics[width=\columnwidth]{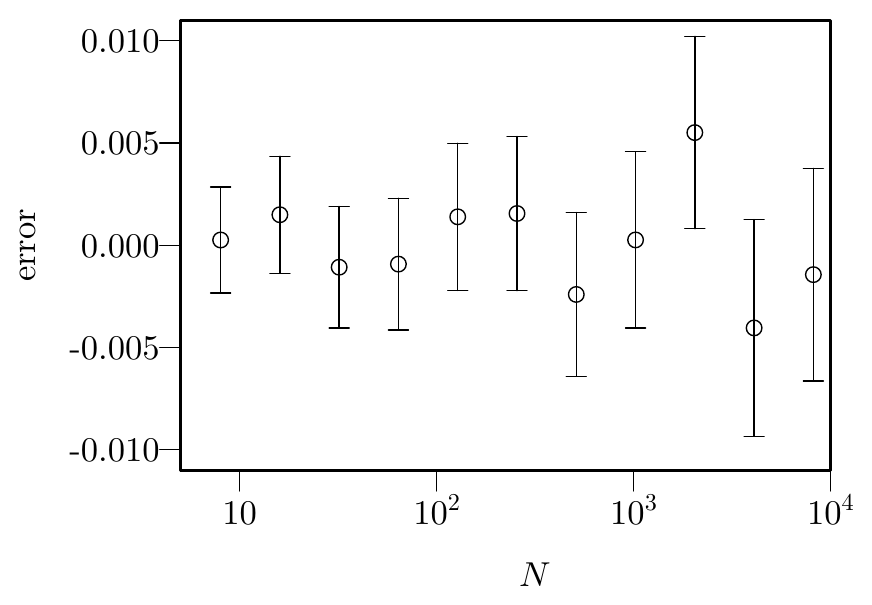}
  \end{center}
  \caption{Same as Fig.~\protect\ref{fig:GaussianFullSample}, but now
    applying the refined sampling strategy where the interparticle
    distance is only evaluated for one randomly selected pair of
    monomers.}
  \label{fig:GaussianRestrictedSample}
\end{figure}

It is also very easy to stochastically generate such a chain using
Gaussian random numbers, based upon the Box-Muller transformation. We
therefore studied chains of length $8 \le N \le 8192$ and sampled
$\avRHinv$ from $10^6$ stochastic realizations. We first calculated
$\RHinv$ in the conventional way by brute-force evaluation of the
double sum. Using the same kind of plot as in
Fig.~\ref{fig:GaussianExact}, the results are indistinguishable from
the exact values. We hence rather show the deviation from the exact
result, using the same normalization as in
Fig.~\ref{fig:GaussianExact} (i.e. we study $\avRHinv$ normalized by
the asymptotic gyration radius of a chain with the same $N$). The
result is shown in Fig.~\ref{fig:GaussianFullSample}. As it should be,
the sampled results are well compatible with the exact values within
error bars.

\begin{figure}
  \begin{center}
    \includegraphics[width=\columnwidth]{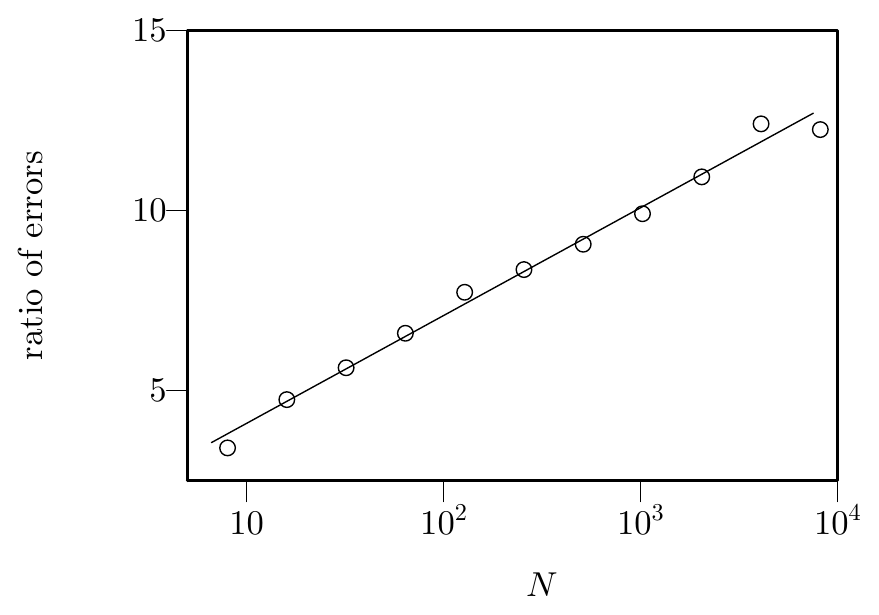}
  \end{center}
  \caption{Ratio of the error bars from
    Figs.~\protect\ref{fig:GaussianRestrictedSample} and
    \protect\ref{fig:GaussianFullSample}, as a function of chain
    length $N$.  As the statistical uncertainty of these data was not
    sampled, we do not show statistical error bars.  The straight line
    is the function $1.1 + 1.3 \log N$.}
  \label{fig:GaussianRatioOfErrorBars}
\end{figure}

Using the same chains, we then sampled $\avRHinv$ by
the ``one pair of monomers per chain'' sampling strategy as outlined
above. As seen in Fig.~\ref{fig:GaussianRestrictedSample}, again the
results are nicely compatible with the exact values within error
bars. The important point to notice is that the latter are only
roughly a factor of 10 larger than in the case of full evaluation,
and this ratio varies only very weakly (possibly logarithmically) with
chain length, as shown in Fig.~\ref{fig:GaussianRatioOfErrorBars}.
This however means quite clearly that the immense computational effort
to evaluate the double sum does not pay off in terms of a
substantially increased statistical accuracy, and that rather the
``one pair of monomers per chain'' method is a much more efficient
overall sampling strategy. One may think of a variant of this scheme,
where one rather picks pairs $(i,j)$ not uniformly, but rather with a
probability $\propto \vert i - j \vert^{-\alpha}$ for some $\alpha$. 
However,
we expect such a change to only slightly improve the statistical
accuracy, compared to the tremendous gain obtained by discarding the
double sum. We hence did not try such a refinement and kept
using simple uniform sampling.

At this point, we wish to remark that it may also be useful to pick
more than just one pair of monomers per chain. This of course
helps to improve the statistical
accuracy somewhat. More importantly, however, this is needed if one is
interested not only in the average value but also in higher moments of
the distribution or in time correlation functions that characterize
the efficiency of the algorithm. Let us discuss this in more detail
for the variance of the inverse hydrodynamic radius. Obviously, we have
\begin{align}
    \text{var}(\RHinv) & = 
  \left\langle \RH^{-2} \right\rangle - \left\langle \RHinv \right\rangle^2
  \\
  \nonumber
  & = 
  \frac{1}{N^4} \sum_{i \ne j} \sum_{k \ne l} \left\{
  \left\langle \frac{1}{r_{ij}} \frac{1}{r_{kl}} \right\rangle -
  \left\langle \frac{1}{r_{ij}} \right\rangle
  \left\langle \frac{1}{r_{kl}} \right\rangle
  \right\}
  \\
  \nonumber
  & = 
  \left(1 - \frac{1}{N} \right)^2
  \left\{
  \left\langle \left[ \frac{1}{r} \right]^2 \right\rangle -
  \left\langle \left[ \frac{1}{r} \right] \right\rangle^2
  \right\} ,
\end{align}
where the last step is performed by using the same trick
as in (\ref{eq:sumtoaverage}) to convert the sum over monomers to an
average.
To sample this by a one-pair-per-chain strategy is 
impossible, however, since the form $\langle [ 1/r ]^2 \rangle$ no
longer permits us to just exchange the averages $\langle \cdots \rangle$
and $[ \cdots ]$. Rather we have
\begin{align}
  \left[ \frac{1}{r} \right]^2
    &= \left[ \left[ \frac{1}{r} \frac{1}{r'} \right] \right] ,
\end{align}
where $[[ \cdots ]]$ is now an average
involving \emph{four} monomers $i, j, k, l$ with $i \ne j$ and $k \ne
l$. To obtain this average, one needs to randomly pick such four
monomers and calculate $r_{ij}^{-1} r_{kl}^{-1}$. This latter average
is again interchangeable with $\langle \cdots \rangle$ and
hence is in accord with
our general strategy. Similar considerations apply for even higher
moments, or time correlation functions. These considerations have
motivated us to run the simulation by not sampling one but rather two
monomer pairs per chain.

In practice, for the main computer experiment of self-avoiding walks,
the observable we sample is
\begin{align}
    \label{eq:q}
    Q &= \frac{1}{2} \left(1 - \frac{1}{N}\right)
    \left( \frac{1}{r} + \frac{1}{r'} \right),
\end{align}
which satisfies $\langle [[ Q ]] \rangle = \langle \RHinv \rangle$.

\section{Details of computer experiment}
\label{sec:details}

We now briefly describe the details of the computer experiment, which
involved the pivot algorithm sampling of self-avoiding walks for
which the number of monomers $N$ varied from 512 to 33554432
($2^{25}$).

The pivot algorithm is ergodic and satisfies the detailed
balance condition~\cite{madras_pivot_1988}, and so samples 
self-avoiding walks uniformly at
random. However, to avoid initialization bias it is necessary to run
the pivot algorithm until the Markov chain is indistinguishably close to
equilibrium.
In each case the seed self-avoiding walk was generated using the
pseudo-dimerize algorithm described in
Ref.~\cite{clisby_efficient_2010}; the system was then equilibrated by
performing approximately $20 N$ successful pivots (no data were
collected during the initialization stage).

Now that an appropriate initial SAW configuration had been generated,
the computer experiment to collect data was begun.
At each time step various observables were sampled: the exact values
for the squared end-to-end distance and the radius of gyration were
used, while the inverse hydrodynamic radius, and the square of the
inverse hydrodynamic radius were estimated using an unbiased
estimator, as described in Sec. \ref{sec:sampling}.

The computer experiment was run for 195 thousand CPU hours on Dell
PowerEdge FC630 machines with Intel Xeon E5-2680 CPUs (these were run
in hyperthreaded mode which gave a modest performance boost;
390 thousand CPU thread hours were used). In total there were
$1.70 \times 10^6$ batches of $10^8$ attempted pivots, and thus there
were a grand total of $1.70 \times 10^{14}$ attempted pivots across all
walk sizes.

We confirmed that the batching method of error estimation was reliably
converging even for the largest values of $N$. This indicates that the
degree of correlation between consecutive batches of $10^8$ pivot
attempts was minimal for each of our global observables $\REsq$,
$\RGsq$, and $\RHinv$, even for the largest size where $N=2^{25}$.

The raw data that have been produced in this way are given in the
tables of Appendix~\ref{sec:data}. We include estimates of the
amplitude ratios $\avREsq / \avRGsq$ and $\avRGsq^{1/2} \avRHinv$ as
they have smaller confidence intervals than might naively be expected
from the estimates of $\avREsq$, $\avRGsq$ and $\avRHinv$ due to
correlations between the observables $\REsq$, $\RGsq$, and $\RHinv$
which reduce the variance of the ratio estimates.

We now briefly consider the properties of our novel Markov chain
sampling method, with a view to gauging the relative effectiveness of
our method for $\RHinv$ versus the observable $\REsq$ which can be
calculated exactly in an efficient manner.

Given an observable $A$ with variance
$\mathrm{var}(A) = \langle A^2 \rangle - \langle A \rangle^2$,
we follow Ref.~\cite{li_critical_1995} and define the autocorrelation function for 
this observable as
\begin{align}
\rho_{AA}(t) &= \frac{\langle A_s A_{s+t} \rangle - \langle A
  \rangle^2}{\mathrm{var}(A)}.
\end{align}

The key quantity which measures the efficiency with which $A$ is sampled
is the integrated autocorrelation time $\tint$, defined as
\begin{align}
\tau_\mathrm{int}(A) &= \frac{1}{2} + \sum_{t=1}^{\infty} \rho_{AA}(t).
\end{align}
$\tint$ may be thought of as the number of Markov chain steps required
before the state is effectively new with respect to the observable $A$.
For a sampling scheme where consecutive estimates are completely
uncorrelated we would have $\tint(A) = 1/2$.
While $\tint$ may well be different for different observables, for
the pivot algorithm we expect that global observables such as $\RGsq$,
$\REsq$, and $\RHinv$ should decorrelate after a constant number of
successful pivots.

We can then calculate an \emph{a priori} estimate of the expected error
on our estimate of the sample mean $\bar{A}$ for $n_{\text{sample}}$ Markov
chain time steps:
\begin{align}
\mathrm{stdev}(\bar{A}) &=
    \left(\frac{2\tau_{\mathrm{int}}(A)\mathrm{var}(A)}
    {n_{\text{sample}}}\right)^{\frac{1}{2}}.
\end{align}

Our goal in performing our Monte Carlo simulation is to estimate
$\langle A \rangle$ as accurately as possible for a given amount of
computer time.  Usually, this entails either finding an observable
$A'$ for which $\langle A' \rangle = \langle A \rangle$ but $\var(A')
< \var(A)$, thus allowing for more efficient sampling (variance
reduction), or finding a Markov chain with an improved move set which
reduces $\tint(A)$, or improving the efficiency of the computer
implementation which allows $n_{\text{sample}}$ to be increased for
the same computational effort.

Our situation is a unique mix of these: We instead estimate an
observable $Q$ from Eq.~\ref{eq:q}
which can be much more efficiently evaluated, thus increasing
$n_{\text{sample}}$, but at the expense of increasing the variance. The
key question is: what is the performance penalty from doing this,
relative to an efficient exact method?

\begin{figure}[tb]
  \begin{center}
    \includegraphics[width=\columnwidth]{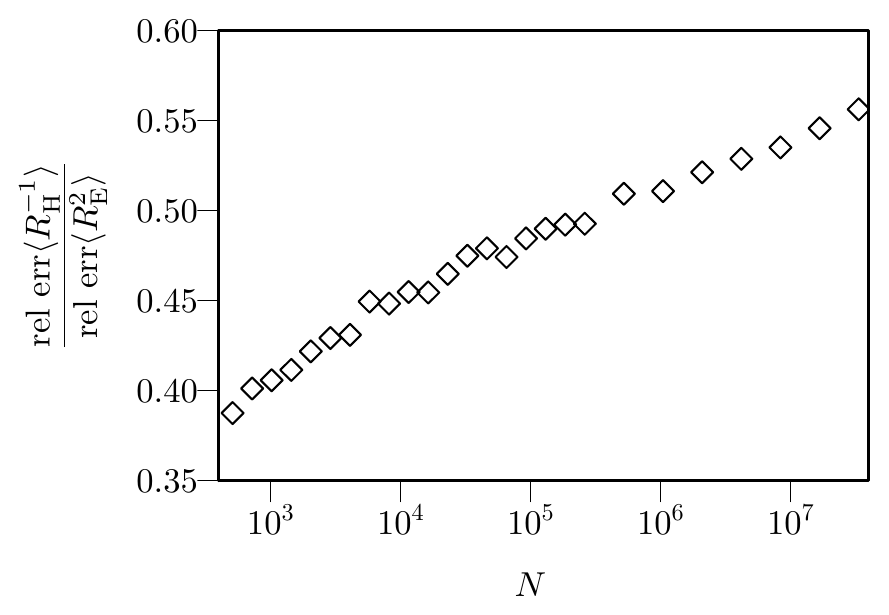}
  \end{center}
  \caption{Plot of the ratio of relative errors for $\avRHinv$ and $\avREsq$.}
  \label{fig:errorratio}
\end{figure}

We examine this question by calculating the ratio of relative errors in
the estimates of $\avRHinv$ and $\avREsq$ which we plot in
Fig.~\ref{fig:errorratio}. There we see that the relative error for 
$\avRHinv$ is substantially below that for $\avREsq$, although the ratio
is growing with $N$, perhaps logarithmically.
This behavior is qualitatively the same as the situation for a Gaussian
chain as shown in Fig.~\ref{fig:GaussianRatioOfErrorBars}. In fact, we
expect that the relative performance penalty should be somewhat less
than that case, because pivot moves are only successful on average once
every $O(N^{p})$
attempts ($p \approx 0.11$ for the simple cubic lattice), and so $Q$ is
sampled on $O(N^p)$ occasions over a time period for which $\RHinv$ remains
frozen.

Thus it seems that the performance penalty is quite modest. Whether
there exist alternatives to the observable $Q$ which could significantly
improve sampling performance is an open research question.

\section{Analysis and Results}
\label{sec:analysis}

\begin{figure}[t]
  \begin{center}
    \includegraphics[width=\columnwidth]{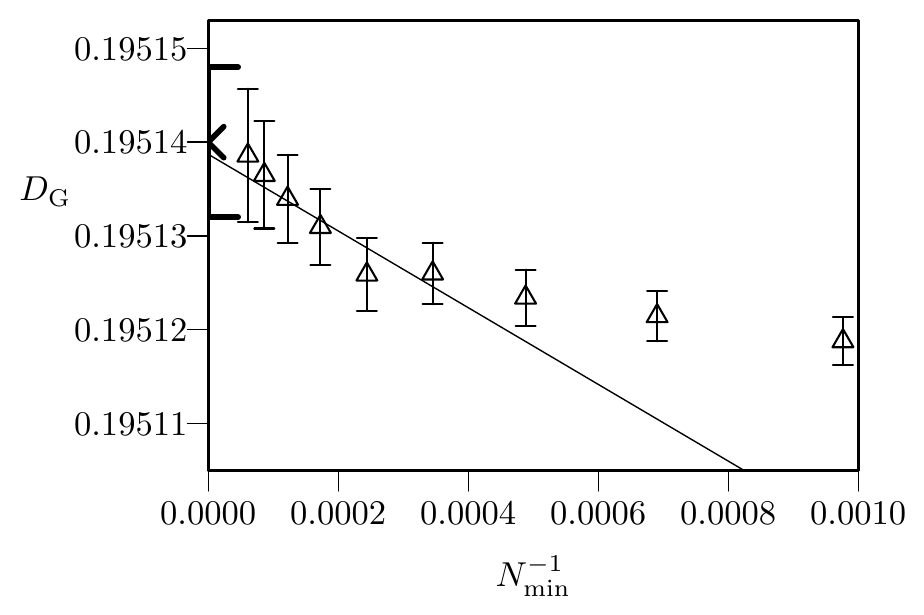}
  \end{center}
  \caption{Systematic variation of the fitted amplitude of $\avRGsq$
    with $\Nmin$. 
    The line of best fit to the final six values is shown, 
    and we plot our best
    estimate from these data of $\DDG = \ourstandardDG$. }
  \label{fig:DG}
\end{figure}

\begin{figure}[tb]
  \begin{center}
    \includegraphics[width=\columnwidth]{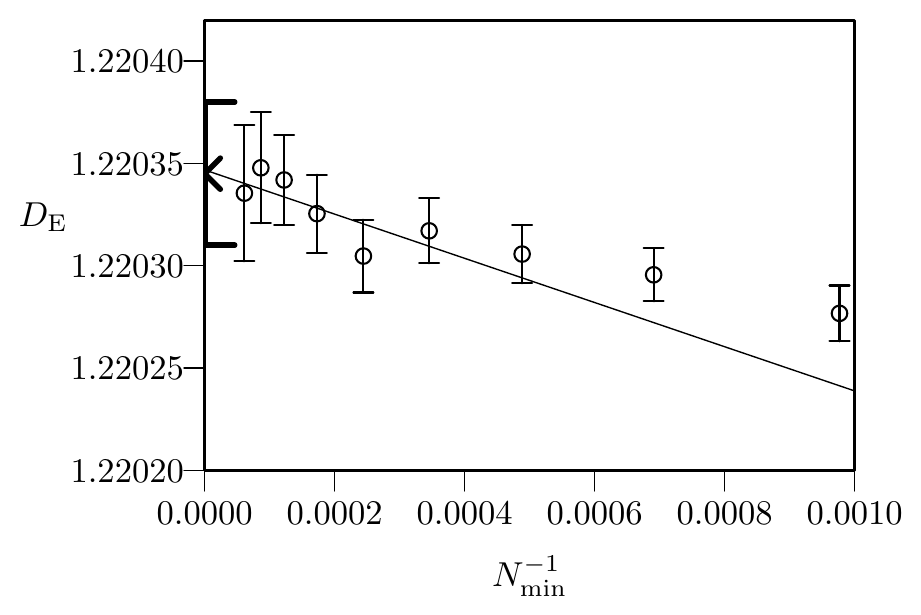}
  \end{center}
  \caption{Systematic variation of the fitted amplitude of $\avREsq$ with $\Nmin$.
    The line of best fit to the final six values is shown, 
    and we plot our best
    estimate from these data of $\DDE = \ourstandardDE$. }
  \label{fig:DE}
\end{figure}

In this section we describe the analysis of data collected in the tables
of Appendix~\ref{sec:data}.  We initially fit the data for standard
observables with a model derived from their expected asymptotic
behavior; this is the conventional method. We then describe a method
which has been used previously for the Ising
model~\cite{Hasenbusch20073dDilutedIsingImprovedObservable,Hasenbusch2010Finitesizescaling},
which eliminates the leading order correction to scaling term and allows
for a much improved estimate for $\nu$.  Next we analyze our data for
the hydrodynamic radius, and present a summary of our results together
with estimates from the literature in Table~\ref{tab:universal}.

We first study the data for $\avRGsq$. Starting from
Eq.~\ref{eq:ExpansionRG}, we apply four-parameter fits to the data,
where $\DDG$, $a_{\mathrm{G}}$, $\nu$, and $\Delta_1$ are considered as
fit parameters, while the higher-order corrections to scaling are
neglected. Because of the large range of chain lengths and the high
resolution accessible to our simulation, these higher-order corrections
cause systematic errors in the fits at a comparable level to the
statistical error. For this reason, we do the fits for various ranges of
chain lengths ($N \ge \Nmin$, where $\Nmin$ is varied systematically).
The effect of the higher-order corrections is then a systematic
dependence of the fit parameters on $\Nmin$. In fact, the deviations for
$\DDG$ and $\nu$ are expected to scale as $\Nmin^{-y}$, where $y$ is the
correction-to-scaling exponent corresponding to the first neglected term
(for a derivation, see Appendix~\ref{sec:fitting}).  In
Eq.~\ref{eq:ExpansionRG} it is believed that there are in fact three
competing next-to-leading correction terms with exponents 1 (analytic),
$2\Delta_1 \approx 1.06$, and $\Delta_2 \approx 1$ ($\Delta_2$ is not
known with any precision).  Assuming a value $y \approx 1$ we thus plot
the estimates for $\DDG$ and $\nu$ as a function of $\Nmin^{-1}$. For
$\avREsq$ we can apply the same analysis to Eq.~\ref{eq:ExpansionRE}.

We perform one further trick to reduce the influence of unfitted
correction to scaling terms and make extrapolation easier. We multiply
our raw data by $1-c/N$, where $c$ is an arbitrary constant chosen to
reduce the curvature observed in fits. Note that this trick does not
change the leading or next-to-leading asymptotic behavior of the
observables, and so if extrapolation is performed carefully this will
not affect our final estimates. We found that a good choice 
for $\avRGsq$ was $c=0.0$, for $\avREsq$ we had $c=0.6$, 
for $\avREsq/\avRGsq$ we had $c=0.2$, for $\avRHinv$ we had $c=-0.2$, 
and for $\avRGsq^{1/2} \avRHinv$ we had $c=-0.5$.

\begin{figure}[t]
  \begin{center}
    \includegraphics[width=\columnwidth]{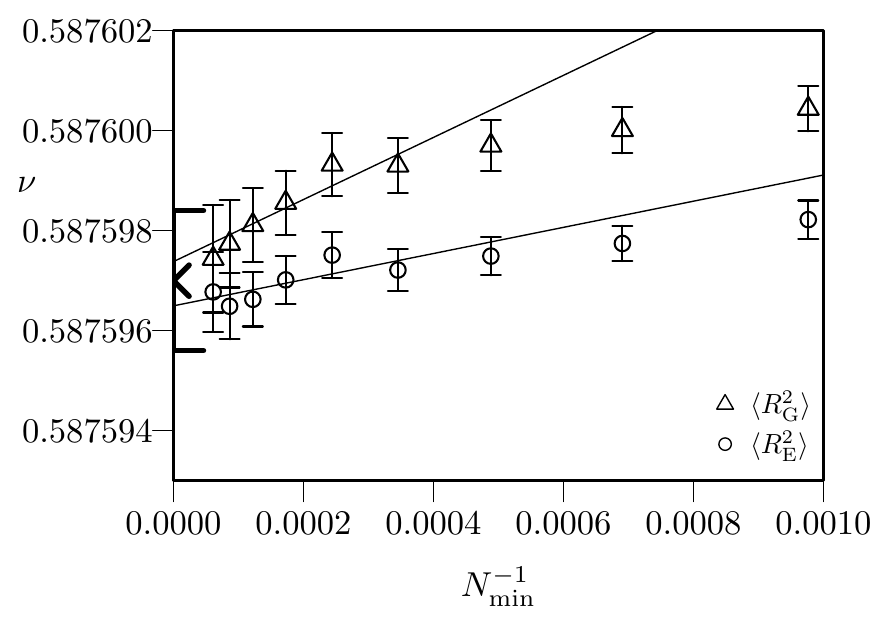}
  \end{center}
  \caption{Systematic variation of the fitted value of $\nu$ with
    $\Nmin$, using both $\avRGsq$ and $\avREsq$ data. 
    The line of best fit to the final six values is shown, 
    and we show our best estimate from these fits of
    $\nu = \ourstandardnu$.}
  \label{fig:nu}
\end{figure}

We plot the resulting estimates in
Figs.~\ref{fig:DG}, \ref{fig:DE}, and \ref{fig:nu}. Note that all error
bars shown are statistical and arise from the fitting procedure. To take
into account the systematic error from corrections to scaling we
extrapolate to the left-hand side of the plots where $\Nmin \rightarrow \infty$. We
choose our final extrapolated value for the parameters by performing
linear fits of subsequent estimates, with an error bar which is
sufficiently large so as to account for both the observed statistical
error and unobserved systematic error which manifests itself in the
plots as non-linear convergence. In the case of Fig.~\ref{fig:nu} we
have the benefit of two observables giving estimates for $\nu$ which
have different unfitted corrections, which increases the reliability of
the extrapolation procedure.

The fit in Fig.~\ref{fig:nu} gives $\nu = \ourstandardnu$ which improves
significantly on the literature, but we can do better as we show later
in this section!  Note that throughout this work we usually report two
significant figures for our confidence intervals. This is not because we
claim that these confidence intervals are so precise, but because
information is lost when only one significant figure is used. For
example, 
confidence intervals of $35 \times 10^{-8}$ and $44 \times 10^{-8}$
would both be reported as a confidence interval of $4 \times 10^{-7}$ if
only one significant figure were used.

\begin{figure}[tb]
  \begin{center}
    \includegraphics[width=\columnwidth]{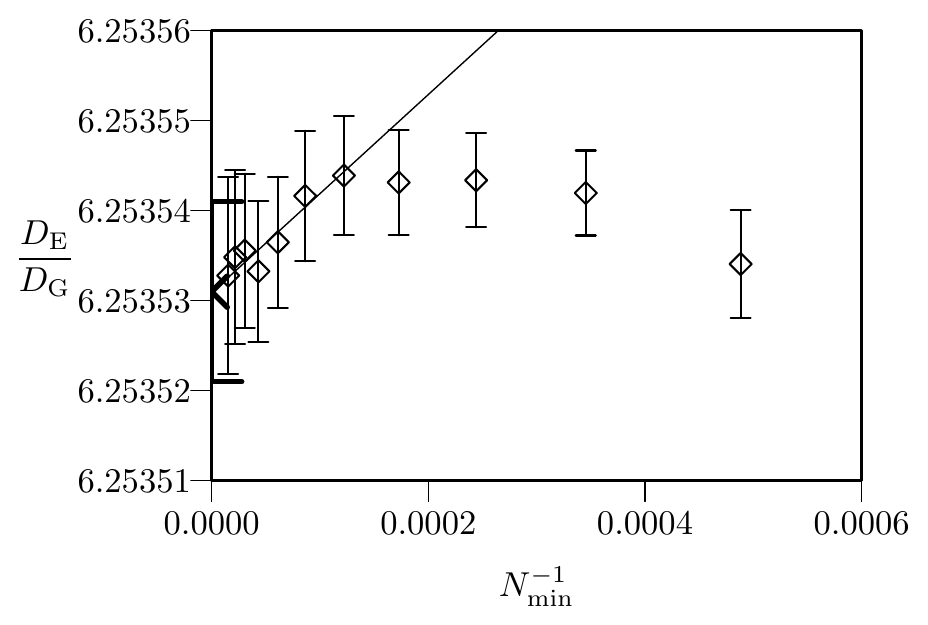}
  \end{center}
  \caption{Systematic variation of our estimates of $\DDE/\DDG$ with $\Nmin$.
    The line of best fit to the final six values is shown, 
    and we show our best estimate from these
    fits of $\DDE/\DDG = \ourrerg$.}
    \label{fig:re2rg2}
\end{figure}

Similarly, we can also study the ratio $\avREsq / \avRGsq$, 
which converges towards the universal amplitude ratio
$\DDE/\DDG$. Taking the ratio reduces the fits from four to three
parameters, as the powers of $N^{2\nu}$ cancel out, and for this reason
the estimate
$\DDE/\DDG$ is more accurate than for the individual amplitudes $\DDE$
and $\DDG$.
The estimated values should again vary systematically like
$\Nmin^{-1}$, and the corresponding plot
is Fig.~\ref{fig:re2rg2}. The universal ratio is therefore
found to take the asymptotic value $\DDE/\DDG = \ourrerg$.

\begin{figure}[bt]
  \begin{center}
    \includegraphics[width=\columnwidth]{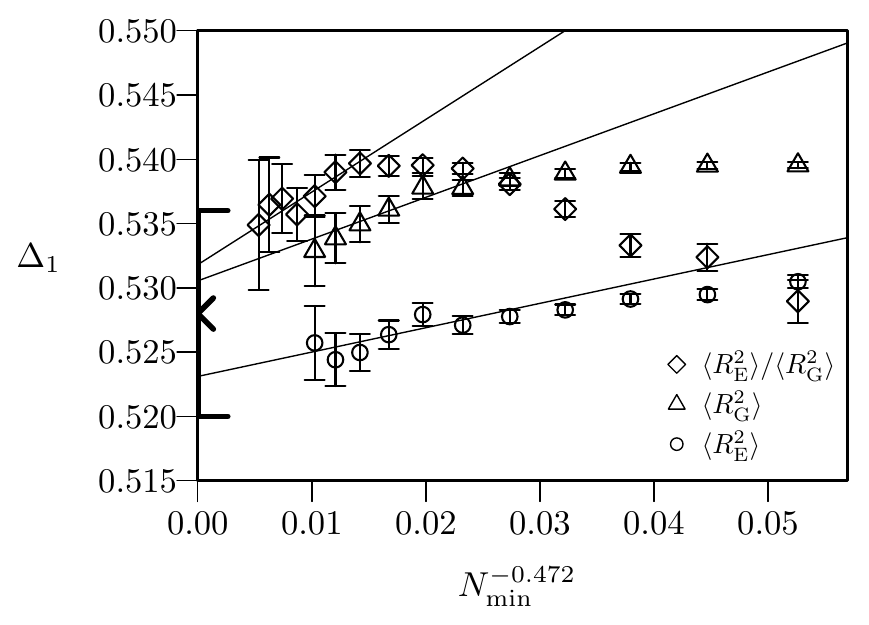}
  \end{center}
  \caption{Systematic variation of the fitted value of $\Delta_1$ with
    $\Nmin$, using data for $\avRGsq$, $\avREsq$ and their ratio.
    The line of best fit to the final six values is shown, 
    and we show our best estimate from these
    fits of $\Delta_1 = \ourdelta$.}
  \label{fig:delta1}
\end{figure}

Finally, we can also use these data to determine $\Delta_1$, whose
value is found to be $\Delta_1 = \ourdelta$. Again taking the next to
leading correction exponent as $-1$, the fitted value
should vary with $\Nmin$ like $\Nmin^{\Delta_1 - 1} \approx \Nmin^{-0.472}$. The
results are shown in Fig.~\ref{fig:delta1}.

\begin{figure}[t]
  \begin{center}
    \includegraphics[width=\columnwidth]{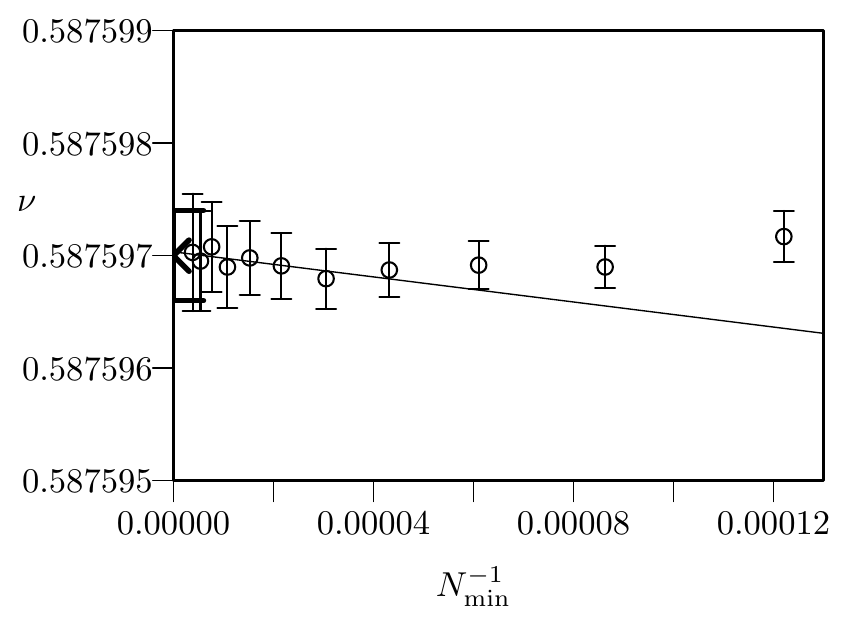}
  \end{center}
  \caption{Systematic variation of the fitted value of $\nu$ with
    $\Nmin$, using data for the improved combination $\avRimp = \avREsq - 4.478 \avRGsq$.
    The line of best fit to the final six values is shown, 
    and we show our best estimate from these fits of
    $\nu = \ournu$.}
  \label{fig:nuimproved}
\end{figure}

\begin{figure}[t]
  \begin{center}
    \includegraphics[width=\columnwidth]{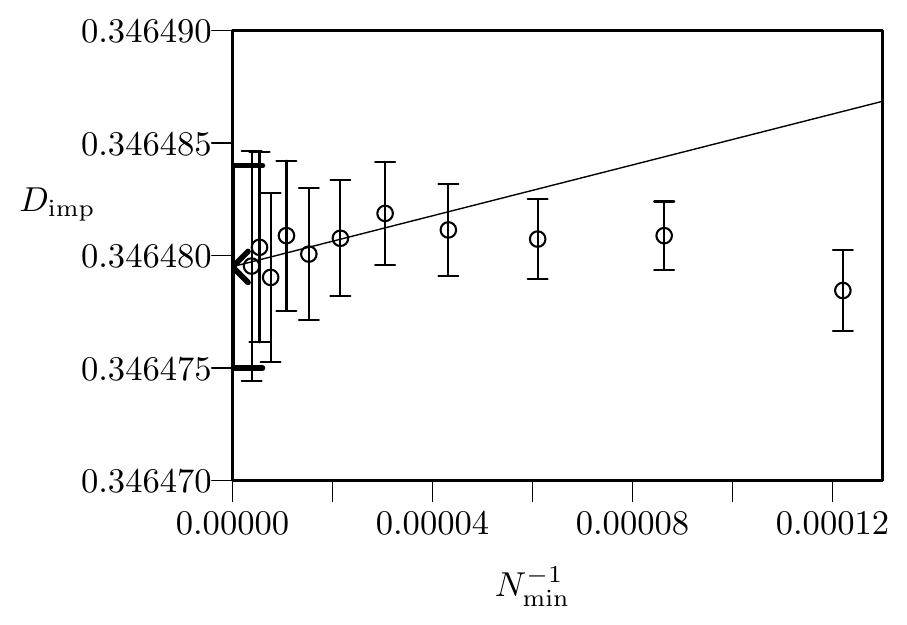}
  \end{center}
  \caption{Systematic variation of the fitted value of $\DDimp = \DDE - 4.478
    \DDG$ with
    $\Nmin$, using data for the improved combination $\avRimp = \avREsq - 4.478 \avRGsq$.
    The line of best fit to the final six values is shown, 
    and we show our best estimate from this
    fit of $\DDimp = \ourDimproved$.}
  \label{fig:Dimproved}
\end{figure}

We now describe a method of analysis which allows us to eliminate
the leading correction to scaling and obtain a much improved estimate
for $\nu$.

It is a standard technique to use improved models for simulations in
statistical mechanics, where typically a parameter is chosen so that the
leading correction to scaling term for all observables is reduced
sufficiently so that their contributions are below the level of
statistical error. For models in the self-avoiding walk universality
class, two such improved models are the Domb-Joyce
model~\cite{caracciolo_polymer_2006} and the
bead model~\cite{kremer_diploma_thesis}.

The basic idea of the method is very simple: instead of attempting to find an improved
model, we find an improved observable instead. 
This technique was previously used for the
three-dimensional dilute Ising
model~\cite{Hasenbusch20073dDilutedIsingImprovedObservable}
and models in the universality class of the three-dimensional Ising
model~\cite{Hasenbusch2010Finitesizescaling}.
Since $\avRGsq$ and
$\avREsq$ are independent measures of the size of a polymer, the
relative size of the leading correction to scaling term for each of
these observables is different. By forming an improved observable 
$\Rimp$ via the linear combination
\begin{align}
    \label{eq:Rimp}
    \Rimp &= \REsq - 4.478 \RGsq,
\end{align}
we find that we are able to reduce the amplitude of the leading
correction to scaling to a level below the statistical noise. 

We are then able to fit $\avRimp$ by the truncated model 
\begin{align}
\avRimp &= \DDimp N^{2\nu}\left(1 +
\frac{\varepsilon}{N^{\Delta_1}}
+ O\left(\frac{1}{N}\right) \right),
\end{align}
where we only fit $\DDimp$ and $\nu$, neglecting the $O(\varepsilon)$
term.
We confirm that this is indeed an excellent model for the data for $\Nmin
\geq 8192$ as the reduced $\chi^2$ of the fits is approximately 1.
By reducing the order of the fits from four
parameters to two, we obtain sensible fits even for $\Nmin$ up to
262144 which are far more accurate than the estimates from fits of
$\avREsq$ and $\avRGsq$. We plot the resulting estimates for $\nu$
against $\Nmin^{-1}$ in Fig.~\ref{fig:nuimproved}, where it can be seen
that convergence in the limit $\Nmin \rightarrow \infty$ is smooth. 

Note that in this case we did not use the
additional trick
of multiplying by $1-c/N$. We have also checked the stability of the
method by varying the constant in Eq.~\ref{eq:Rimp}, and find that within
the interval $(4.473,4.483)$ the plot in Fig.~\ref{fig:nuimproved} is quite
linear and can be extrapolated easily.

Note the substantial decrease in range and domain for the plots from the
standard approach in Fig.~\ref{fig:nu} as compared to the new approach
in Fig.~\ref{fig:nuimproved}. 
Purely from this novel method of analysis we have
managed to decrease the error by more than a factor of three, from 
$14\times 10^{-7}$ to $4 \times 10^{-7}$. Our central estimate has not
changed, and our final estimate is $\nu = \ournu$.

We now perform one final trick to obtain improved estimates for $\DDE$
and $\DDG$. We first plot the estimates for $\DDimp$ obtained from our
two-parameter fits in Fig.~\ref{fig:Dimproved}. We then use the fact
that our estimate of $\DDE/\DDG$ is more accurate than the estimates of
$\DDE$ and $\DDG$ individually, and form the combinations:
\begin{align}
    \DDE &= \frac{\DDimp}{1 - 4.478 \DDG/\DDE}, \\
    \DDG &= \frac{\DDimp}{\DDE/\DDG - 4.478}.
\end{align}
We combine the errors from $\DDimp$ and $\DDE/\DDG$ as if they were
independent, and obtain the improved estimates
$\DDG = \ourDG$ and $\DDE = \ourDE$.

\begin{figure}[b]
  \begin{center}
    \includegraphics[width=\columnwidth]{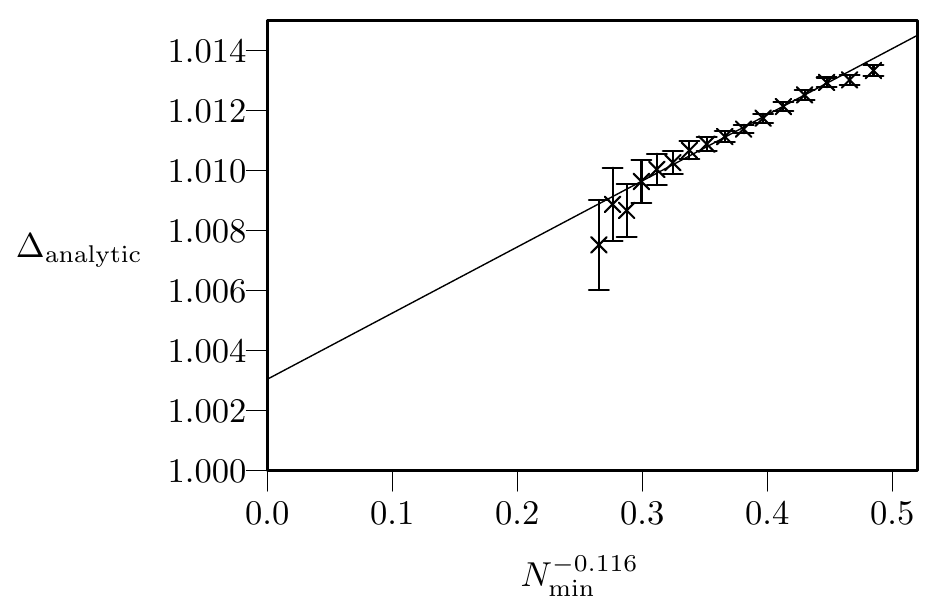}
  \end{center}
  \caption{Systematic variation of the fitted value of $\Delta_{a}$ with
    $\Nmin$, using data for $\avRHinv$.}
  \label{fig:rhone}
\end{figure}

We now turn to the $\RHinv$ data, where Eq.~\ref{eq:ExpansionRH} applies.
Again, we start with a four-parameter fit, where we take the leading
order into account, plus the dominant correction to scaling. The
latter should be the analytic term, which is absent for $\avRGsq$ and $\avREsq$. If only those two terms are
present, the fit function can be written as
\begin{align}
  \label{eq:rhone}
    \avRHinv &= \DDH N^{-\nu} + E_{\mathrm{H}} N^{-\Delta_{a}},
\end{align}
where the analytic value $\Delta_{a}$ is one. This contribution is
difficult to distinguish from the next-order contribution, which
scales as $N^{- (\nu + \Delta_1)} \approx N^{-1.116}$, where the
exponent is only slightly different. However, our data are accurate
enough that this is actually possible. We therefore apply a
four-parameter fit to the data according to Eq.~\ref{eq:rhone}, where
$\Delta_{a}$ is left as a fit parameter. Using the results of
Appendix~\ref{sec:fitting}, these data should then vary with $\Nmin$
according to $\Delta_{a} \propto \Nmin^{\Delta_a - \nu - \Delta_1} =
\Nmin^{-0.116}$. As seen in Fig.~\ref{fig:rhone}, they nicely
extrapolate to $\Delta_{a} \approx 1$, with a value that is clearly
distinguishable from the next order ($1.116$).

\begin{figure}[b]
  \begin{center}
    \includegraphics[width=\columnwidth]{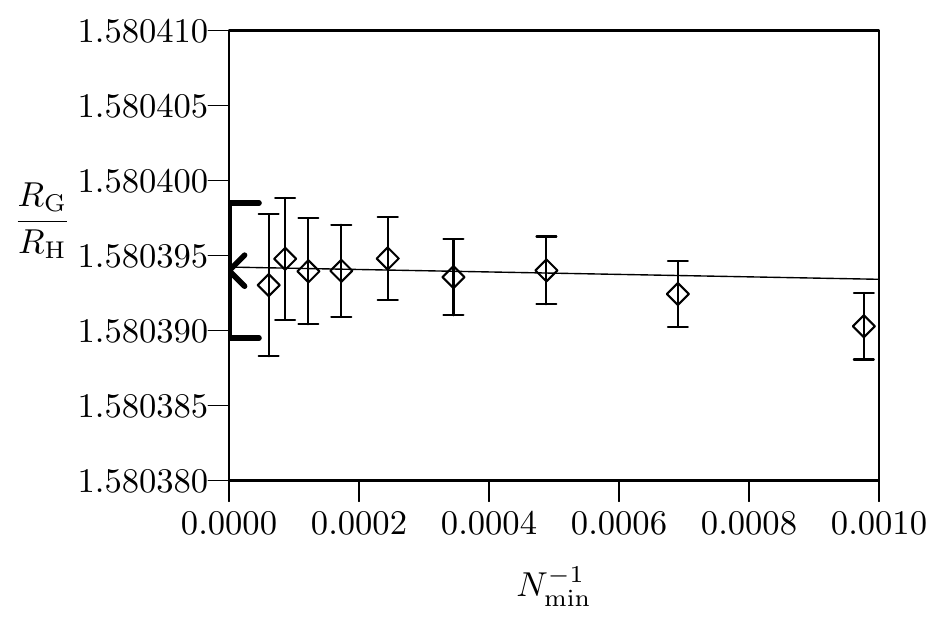}
  \end{center}
  \caption{Systematic variation of the fitted value of $\RG / \RH$ with
    $\Nmin$.
    The line of best fit to the final six values is shown, 
    and we plot our best
    estimate from these data of $\RG/\RH = \ourrgrh$. }
  \label{fig:rgrh}
\end{figure}

Finally, we focus on the universal amplitude ratio $\RG/\RH$, which
was the original motivation to perform the present study. This can
be written as
\begin{align}
  \label{eq:rgrhratio}
    \avRGsq^{1/2} \avRHinv &= \frac{\RG}{\RH} + B N^{-(1 - \nu)} + C N^{-\Delta_1} + \cdots ,
\end{align}
where the omitted leading-order correction is $O(N^{-1})$.  We now use
the value for $\nu$ as obtained from the $\avRGsq$ and $\avREsq$ data,
and treat the parameters $\RG/\RH$, $B$, $C$, and $\Delta_1$ in
Eq.~\ref{eq:rgrhratio} as fit parameters in a four-parameter fit.  The
parameter $\RG/\RH$ should then vary linearly with $\Nmin^{-1}$. The
data are shown in Fig.~\ref{fig:rgrh} and give rise to an estimate for
the universal amplitude ratio of $\RG/\RH = \ourrgrh$.

Table \ref{tab:universal} summarizes our results, 
with a comparison with previous
results from Monte Carlo, series expansion, field theoretic, and
conformal bootstrap methods.  We wish to highlight the recent conformal
bootstrap estimate of $\nu =
0.58775(83)$~\cite{Shimada2015NuForSAWandIsingArxiv} as this approach
shows a great deal of promise. The method has been spectacularly
successful for the three-dimensional Ising model giving $\nu =
0.6299748(40)$~\cite{Kos2016ConformalBootstrapIsingAndONmodels}; in this
case it is far superior to Monte Carlo methods.

\setlength{\abovecaptionskip}{5pt}
\begin{table}[!ht]
\begin{ruledtabular}
\begin{tabular}{lllll}
\multicolumn{1}{c}{Source\footnote{Abbreviations: MC, Monte
    Carlo; CB, conformal bootstrap; FT, field theory; 
    MCRG, Monte Carlo renormalization group.}} 
    & \multicolumn{1}{c}{$\nu$} & \multicolumn{1}{c}{$\Delta_1$} &
    \multicolumn{1}{c}{$\RG/\RH$}
   \\ \hline
    Present work & \ournu & \ourdelta &  \ourrgrh \\
    \cite{Shimada2015NuForSAWandIsingArxiv} CB & 0.58775(83) & & \\
    \cite{Schram2011ExactEnumerationsSelfAvoidingWalks} Series & 0.58772(17) & & \\
    \cite{clisby_accurate_2010} MC & 0.587597(7) & 0.528(12) & \\
\cite{Clisby2007Selfavoidingwalk}\footnote{Using Eqs. (74) and (75) of
    Ref.~\cite{Clisby2007Selfavoidingwalk} with $0.516 \leq \Delta_1 \leq 0.54$.} Series  &   0.58774(22) & &\\
    \cite{caracciolo_polymer_2006} MC & & & 1.581(1) \\
    \cite{dunweg_corrections_2002} MC & & & 1.591(7) \\
\cite{Prellberg2001Scalingselfavoiding} MC & 0.5874(2)    & & \\
\cite{MacDonald2000Selfavoidingwalks}\footnote{No error
  estimates were made in Ref.~\cite{MacDonald2000Selfavoidingwalks}, but estimates for
  $\nu$ were in the 
range $0.5870 \leq \nu \leq 0.5881$.} Series & 0.58755(55)  & &\\
\cite{Guida1998CriticalexponentsN} FT $d=3$& 0.5882(11)  & 0.478(10)   &  \\
\cite{Guida1998CriticalexponentsN} FT $\epsilon$ bc & 0.5878(11)& 0.486(16)  &   \\
\cite{Belohorec1997Renormalizationgroupcalculation} MCRG & 0.58756(5)   & 0.5310(33)   & \\
\cite{li_critical_1995} MC   &  0.5877(6) & 0.56(3) & \\
    \cite{schafer_internal_1986} FT & & & $\approx$ 1.595 \\
\end{tabular}
\end{ruledtabular}
    \caption{ 
    Summary of estimates of $\nu$, $\Delta_1$, and $\RG/\RH$. 
    In addition we
    have $\DDE/\DDG = \ourrerg$ (c.f.
    6.2537(18)~\cite{clisby_accurate_2010}),
    $\DDG = \ourDG$ (c.f. 0.19514(4)~\cite{clisby_accurate_2010}),
    and $\DDE = \ourDE$ (c.f. 1.22035(25)~\cite{clisby_accurate_2010}).
    Note that results in the table are
    listed in reverse chronological order, i.e. the most recently
    published work is at the top.}
    \label{tab:universal}
\end{table}
\setlength{\abovecaptionskip}{-10pt}

\section{Conclusion}
\label{sec:conclusion}

The combination of the pivot algorithm and the SAW-tree data structure
of Refs.~\cite{clisby_accurate_2010,clisby_efficient_2010} provides an
extremely efficient method to obtain the properties of long SAWs with
high accuracy. The SAW-tree allows for the efficient computation of
obervables such as $\RGsq$ and $\REsq$, but not for other observables
such as $\RHinv$, which leads to a unique problem: How to efficiently
sample an observable whose calculation would dominate the runtime of
the Markov chain sampling algorithm?  The key insight is that the
observable does not need to be calculated exactly in order to obtain
accurate estimates, instead we only need to find an unbiased estimator
of the observable which can be calculated efficiently and which has
moderate variance.

Starting from the observation that a large class of observables can be
written as the sum of $n$-body terms involving $n$ monomers, where this
series typically stops at low (and in most cases at second) order, we
propose a double sampling scheme, where not only the chain conformations
are generated at random, but also the monomers that contribute to the
$n$-body interactions are picked at random, such that this evaluation
involving just a few monomers replaces an exhaustive sum over all sets
of $n$ monomers. This leads to an efficient Monte Carlo sampling for
many observables, and the present work demonstrates its usefulness by
applying it to the problem of sampling the hydrodynamic radius of
three-dimensional SAWs. Using this technique we estimated with high
accuracy the universal amplitude ratio $\RG/\RH = \ourrgrh$, and
discerned the competing corrections to scaling for $\avRHinv$. 
Finally, we have constructed an improved observable for which the
leading correction to scaling has negligible amplitude, and used it
to obtain an improved estimate for the Flory exponent of $\nu = \ournu$.

\begin{acknowledgments}
  N.C. acknowledges funding by the Australian Research Council under
  the Future Fellowship scheme (project number FT130100972) and
    Discovery scheme (project
  number DP140101110). B.D. acknowledges hospitality of The
  University of Melbourne during two visits where some of this work
  was conducted.
\end{acknowledgments}

\appendix

\section{Fitting strategy}
\label{sec:fitting}

Here we will describe some of the details of the fitting procedure used in the
main text. The description will be quite general, but we will refer to
specific examples from the analysis section. 

Suppose we are interested in a certain observable, e.g. $\avRGsq$, as
a function of chain length $N$. Let us denote this observable as
$R(N)$. We expect that for $R(N)$ there exists an infinite asymptotic
expansion in $N$:
\begin{align}
  \label{eq:DataFunction}
    R(N) &= \sum_{i = 1}^{\infty} \alpha_i N^{x_i} \\
 &= \sum_{i = 1}^{M} \alpha_i N^{x_i} + O\left(N^{x_{M+1}}\right),
\end{align}
where $x_1 > x_2 > \cdots > x_{M} > x_{M + 1} > \cdots$, such that
$x_1$ describes the leading asymptotic power-law dependence, while the
exponents $x_2, x_3, \cdots$ correspond to the corrections to
scaling. The parameters $\alpha_i$ are the corresponding amplitudes.
Note that the procedure described here can be straightforwardly adapted to
observables with different asymptotic behavior, e.g. exponential
growth with power law corrections.

Now we perform a computer experiment which gives us $R(N)$ for
certain values of $N$. For an enumeration study this information would
be exact but typically involve quite small $N$, while for a Monte Carlo
computer experiment there would be sampling error associated with these
values but one could reach large values of $N$ and reduce the influence
of corrections to scaling. 

Our principal goal in performing the computer experiment is to estimate
some of the quantities associated with this asymptotic expansion such as
the leading exponent $x_1$, the leading-order correction to scaling
exponent $x_2$, and the leading amplitude $\alpha_1$.

We obtain estimates by performing non-linear fits of our data using
Eq.~\ref{eq:DataFunction} by appropriately truncating the expansion
after $M$ terms.  We cannot perform a fit with an arbitrarily large
number of terms, as we only have data over a finite range for $N$.
There may also be asymptotic corrections with comparable exponents
which makes it extremely difficult to reliably distinguish between
them, and for Monte Carlo there is statistical error on $R(N)$.  Each
of these factors is relevant in our case: We have data up to $N =
2^{25}$, our data has statistical error, and next-to-leading
corrections to scaling have comparable exponents which are all around
1: $2\Delta_1 \approx 1.06$, $\Delta_2 \approx 1$, and 1. So, in
practice we can only make reliable fits of the leading correction to
scaling.  It is possible to fit the three competing next-to-leading
corrections with a single ``effective'' term with exponent
approximately one, but it is difficult to see how to sensibly
interpret such a procedure.

In the general case, we attempt to simultaneously adjust all
amplitudes $\alpha_i$ and all exponents $x_i$ by the direct
application of a $2M$-parameter nonlinear fit routine. If we include
in the fit $2M$ data points for $R(N)$ then the fitted function $S(N)$
will be exact at those points, but more frequently we perform a
non-linear weighted least squares fit (weighting appropriately by the
statistical error in our estimates of $R(N)$) and so $S(N)$ will
instead be an approximation. Regardless, by design we have
$S(N) \approx R(N)$, where we are careful to ensure that we can
meaningfully interpret the fit by confirming that the model is
appropriate and the reduced $\chi^2$ value is approximately one. If
the model \emph{is} appropriate then $S(N)$ will be the same as $R(N)$
at the data points to within statistical accuracy, and so $S(N) = R(N)
+ \Delta R(N)$ where $\Delta R(N)$ is of the same order as the
statistical accuracy of our estimate.

The truncation will result in somewhat distorted values for the
amplitudes and exponents in the truncated model. We denote these
errors as $\Delta \alpha_i$ for the amplitudes and $\Delta x_i$ for
the exponents:
\begin{align}
  \label{eq:FitFunction}
    S(N) &= \sum_{i = 1}^M \left(\alpha_i + \Delta\alpha_i \right)
    N^{x_i + \Delta x_i} . 
\end{align}
But we have 
\begin{align}
    S(N) &= R(N) + \Delta R(N) , \\
    \sum_{i = 1}^M \left(\alpha_i + \Delta\alpha_i \right)
    N^{x_i + \Delta x_i}
 &= \sum_{i = 1}^{M} \alpha_i N^{x_i} + O\left(N^{x_{M+1}}\right)
 + \Delta R(N) .
    \label{eq:exactasymptoticrelation}
\end{align}
We restrict attention only to rather large values of $N$, where the
truncated model accurately fits the data, and so
$\vert \Delta \alpha_i \vert \ll \vert \alpha_i \vert$ and
$\vert \Delta x_i \vert \ll \vert x_i \vert$.  In addition, we can
expect that the neglected terms represented by $O(N^{x_{M+1}})$ are
small, and dominated by the first neglected correction to scaling
corresponding to exponent $x_{M+1}$.  In this limit, we may linearize
Eq.~\ref{eq:exactasymptoticrelation} around $\alpha_i$ and $x_i$,
\begin{align}
    \sum_{i = 1}^M \left(\Delta \alpha_i N^{x_i} + \alpha_i
    \Delta x_i
    N^{x_i} \log N \right)
    &= O\left(N^{x_{M+1}}\right) + \Delta R(N).
    \label{eq:approximateasymptoticrelation}
\end{align}

We now perform fits according to Eq.~\ref{eq:FitFunction} in an
interval $N \ge \Nmin$, where $\Nmin$ is systematically varied but
where it must be sufficiently large that the truncated model is
accurate. The errors in estimates depend on this choice $\Nmin$, and
so $\Delta \alpha_i$ and $\Delta x_i$ should be understood to be
implicit functions of $\Nmin$.
Eq.~\ref{eq:approximateasymptoticrelation} is valid for any value of
$N$ in the fitting range, and therefore is valid for $\Nmin$:
\begin{align}
    \sum_{i = 1}^M \left(\Delta \alpha_i \Nmin^{x_i} + \alpha_i
    \Delta x_i
    \Nmin^{x_i} \log \Nmin \right)
    &= O\left(\Nmin^{x_{M+1}}\right) + \Delta R(\Nmin).
    \label{eq:nminasymptoticrelation}
\end{align}

Neglecting logarithmic corrections, and assuming that all error terms on
the left-hand side of Eq.~\ref{eq:nminasymptoticrelation} are of the same order as
the right-hand side, we thus find for the error in the exponents that
\begin{align}
    \Delta x_i &\propto \Nmin^{- (x_i - x_{M+1})}
    + \Nmin^{-x_i} \Delta R(\Nmin),
\end{align}
and similarly for the amplitudes
\begin{align}
    \Delta \alpha_i &\propto \Nmin^{- (x_i - x_{M+1})} + \Nmin^{-x_i}
    \Delta R(\Nmin).
\end{align}
How are we to interpret these expressions, and use them to obtain the most
accurate estimates of $\alpha_i$ and $x_i$ possible? Firstly, note that
$\Delta R(\Nmin)$ is the statistical error, and is a known quantity. The
corresponding statistical error in the estimates for $\alpha_i$ and
$x_i$ are of order $\Nmin^{-x_i} \Delta R(\Nmin)$. 
Typically, we expect that
the statistical errors will increase as $\Nmin$ increases, but the rate
of increase will be smallest 
for the leading term with $i=1$.
In contrast, the systematic errors, of order 
$\Nmin^{- (x_i - x_{M+1})}$ (neglecting logarithmic factors) are
unknown, and decay with
increasing $\Nmin$. This decay is most rapid for the leading term.
By definition, the systematic error from truncation is not fitted, and
so the only way which it can be accounted for in the analysis is to
extrapolate to $\Nmin \rightarrow \infty$ where this error vanishes.
Now, we expect that for sufficiently large $\Nmin$, a plot of $\alpha_i$
and $x_i$ against 
$\Nmin^{- (x_i - x_{M+1})}$ would be linear.
If we have an idea of the value of $x_{M+1}$ -- even if we do not know
it exactly -- plotting our estimates in this way can greatly facilitate
extrapolation.
These observations are the motivation for the various power laws
appearing in plots in the main text.
Then, to interpret these fits requires judgment to decide when $\Nmin$
is sufficiently large that a reliable extrapolation can be made, but as
small as possible so as to reduce statistical error.

Interpretation of the fits is a balancing act between systematic error
and statistical error. Acquiring more data at large values of $N$ may
reduce systematic error at the expense of increasing statistical
error.  One of us (N.C.) is perennially surprised at how subtle the
interpretation of such fits is: In principle, being able to perform
accurate computer experiments for extremely large systems should make
it possible to reduce the influence of corrections to scaling until
they are negligible, but what happens in practice is that the
extremely accurate values make it necessary to incorporate the
leading-order correction to scaling even for $N$ of the order of tens
of millions, and in order to get a good handle on this term it is
necessary to perform computer experiments for $N$ of the order of tens
of thousands, where poorly controlled next-to-leading corrections make
things extremely difficult! One circumstance where
this trap has been avoided is the calculation of the growth
constant $\mu$ for SAWs in
Ref.~\cite{Clisby2013ConnectiveConstant}, but this relies on
the fact that the asymptotic corrections for $\mu$ are smaller than for
critical exponents. 

\section{Monte Carlo data}
\label{sec:data}

The global observables $\RGsq$, $\REsq$, and $\RHinv$ are correlated;
therefore calculating ratios may be viewed as form of variance
reduction. Hence we report the ratios as well.


\begin{center}
\begin{tabular}{rll}
\hline
    $N$ & \multicolumn{1}{c}{$\avREsq$} & \multicolumn{1}{c}{$\avRGsq$}
    \bstrut \\
\hline
512 & 1.8336722(58)\e{3} & 2.9152213(82)\e{2} \\
724 & 2.7631843(94)\e{3} & 4.396899(14)\e{2} \\
1024 & 4.1626998(41)\e{3} & 6.6290075(60)\e{2} \\
1448 & 6.2667402(69)\e{3} & 9.986311(10)\e{2} \\
2048 & 9.433354(11)\e{3} & 1.5040985(16)\e{3} \\
2896 & 1.4192522(18)\e{4} & 2.2640087(26)\e{3} \\
4096 & 2.1353085(28)\e{4} & 3.4076501(41)\e{3} \\
5792 & 3.2112468(46)\e{4} & 5.1264340(69)\e{3} \\
8192 & 4.8297971(73)\e{4} & 7.712466(11)\e{3} \\
11584 & 7.261391(12)\e{4} & 1.1598097(18)\e{4} \\
16384 & 1.0918781(19)\e{5} & 1.7443237(29)\e{4} \\
23168 & 1.6412837(31)\e{5} & 2.6224555(48)\e{4} \\
32768 & 2.4675807(49)\e{5} & 3.9432498(75)\e{4} \\
46336 & 3.7087199(80)\e{5} & 5.927288(12)\e{4} \\
65536 & 5.575269(13)\e{5} & 8.911266(20)\e{4} \\
92672 & 8.378786(18)\e{5} & 1.3393305(27)\e{5} \\
131072 & 1.2594736(32)\e{6} & 2.0133731(50)\e{5} \\
185344 & 1.8926972(46)\e{6} & 3.0258005(71)\e{5} \\
262144 & 2.8449071(51)\e{6} & 4.5482716(79)\e{5} \\
524288 & 6.425547(21)\e{6} & 1.0273486(32)\e{6} \\
1048576 & 1.4512152(53)\e{7} & 2.3203899(83)\e{6} \\
2097152 & 3.277454(13)\e{7} & 5.240600(21)\e{6} \\
4194304 & 7.401657(33)\e{7} & 1.1835309(52)\e{7} \\
8388608 & 1.6715288(79)\e{8} & 2.672847(13)\e{7} \\
16777216 & 3.774819(19)\e{8} & 6.036144(31)\e{7} \\
33554432 & 8.524591(30)\e{8} & 1.3631415(48)\e{8} \\
\hline
\end{tabular}
\end{center}

\newpage

\begin{center}
\begin{tabular}{rll}
\hline
    $N$ & \multicolumn{1}{c}{$\avRHinv$} & 
    \multicolumn{1}{c}{$\langle \RH^{-2} \rangle$} 
    \bstrut \\
\hline
512 & 8.400655(10)\e{-2} & 7.124977(20)\e{-3} \\
724 & 6.9369818(95)\e{-2} & 4.858005(16)\e{-3} \\
1024 & 5.7174946(23)\e{-2} & 3.2998280(32)\e{-3} \\
1448 & 4.7059008(21)\e{-2} & 2.2352724(25)\e{-3} \\
2048 & 3.8678939(19)\e{-2} & 1.5099458(18)\e{-3} \\
2896 & 3.1760618(17)\e{-2} & 1.0180323(14)\e{-3} \\
4096 & 2.6052526(15)\e{-2} & 6.849483(10)\e{-4} \\
5792 & 2.1356062(14)\e{-2} & 4.6023215(78)\e{-4} \\
8192 & 1.7492303(12)\e{-2} & 3.0875098(56)\e{-4} \\
11584 & 1.4321075(11)\e{-2} & 2.0694197(42)\e{-4} \\
16384 & 1.17175446(93)\e{-2} & 1.3853382(31)\e{-4} \\
23168 & 9.5844326(85)\e{-3} & 9.268299(23)\e{-5} \\
32768 & 7.8358154(73)\e{-3} & 6.194747(17)\e{-5} \\
46336 & 6.4050264(66)\e{-3} & 4.138895(13)\e{-5} \\
65536 & 5.2334061(57)\e{-3} & 2.7631557(94)\e{-5} \\
92672 & 4.2756617(44)\e{-3} & 1.8443015(61)\e{-5} \\
131072 & 3.4920527(44)\e{-3} & 1.2302115(50)\e{-5} \\
185344 & 2.8519139(34)\e{-3} & 8.205089(33)\e{-6} \\
262144 & 2.3284895(21)\e{-3} & 5.469584(17)\e{-6} \\
524288 & 1.5518439(26)\e{-3} & 2.429360(14)\e{-6} \\
1048576 & 1.0338339(19)\e{-3} & 1.0781790(76)\e{-6} \\
2097152 & 6.885461(14)\e{-4} & 4.782487(40)\e{-7} \\
4194304 & 4.584827(11)\e{-4} & 2.120460(20)\e{-7} \\
8388608 & 3.0523993(78)\e{-4} & 9.39847(10)\e{-8} \\
16777216 & 2.0319314(57)\e{-4} & 4.164729(53)\e{-8} \\
33554432 & 1.3525101(26)\e{-4} & 1.845312(22)\e{-8} \\
\hline
\end{tabular}
\end{center}

\newpage

\begin{center}
\begin{tabular}{rll}
\hline
    $N$ & \multicolumn{1}{c}{$\avREsq/\avRGsq$} &
    \multicolumn{1}{c}{$\avRGsq^{1/2} \avRHinv$} 
    \bstrut \\
\hline
512 & 6.289993(10) & 1.4343295(15) \\
724 & 6.284393(11) & 1.4546008(17) \\
1024 & 6.2795219(32) & 1.47207524(52) \\
1448 & 6.2753307(35) & 1.48711752(57) \\
2048 & 6.2717661(37) & 1.50007398(61) \\
2896 & 6.2687579(39) & 1.51122100(68) \\
4096 & 6.2662199(42) & 1.52081824(73) \\
5792 & 6.2640947(45) & 1.52907518(80) \\
8192 & 6.2623255(48) & 1.53618530(86) \\
11584 & 6.2608471(51) & 1.54230033(94) \\
16384 & 6.2596072(54) & 1.5475694(10) \\
23168 & 6.2585759(59) & 1.5521027(11) \\
32768 & 6.2577340(62) & 1.5560062(12) \\
46336 & 6.2570265(68) & 1.5593691(13) \\
65536 & 6.2564277(71) & 1.5622629(14) \\
92672 & 6.2559513(66) & 1.5647580(13) \\
131072 & 6.2555401(80) & 1.5669058(16) \\
185344 & 6.2551952(75) & 1.5687600(15) \\
262144 & 6.2549194(55) & 1.5703535(11) \\
524288 & 6.254495(10) & 1.5729210(21) \\
1048576 & 6.254187(11) & 1.5748209(23) \\
2097152 & 6.253967(12) & 1.5762441(26) \\
4194304 & 6.253877(13) & 1.5772942(29) \\
8388608 & 6.253739(15) & 1.5780775(32) \\
16777216 & 6.253693(16) & 1.5786606(35) \\
33554432 & 6.253636(10) & 1.5791045(24) \\
\hline
\end{tabular}
\end{center}


\end{document}